\documentclass[aps,prl,twocolumn,groupedaddress,floatfix]{revtex4-2}

\usepackage{amsmath}
\usepackage{bm}
\usepackage{braket}
\usepackage{txfonts}
\usepackage{color}
\usepackage[dvipdfmx]{graphicx}
\begin{document}


\title{Presence of x-ray magnetic circular dichroism effect \\
for zero-magnetization state in antiferromagnetism}


\author{Norimasa Sasabe$^{1}$ \thanks{sasabe@spring8.or.jp}, Motoi Kimata$^2$, and Tetsuya Nakamura$^{1,2,3}$ }
\affiliation{$^1$Japan Synchrotron Radiation Research Institute, SPring-8 Kouto, Sayo, Hyogo 679-5198, Japan  \\
$^2$Institute for Materials Research, Tohoku University, Sendai, Miyagi 980-8577, Japan \\
$^3$International Center for Synchrotron Radiation Innovation Smart, Tohoku
University, Sendai 980-8577, Japan
}


\date{\today}

\begin{abstract}
X-ray magnetic circular dichroism (XMCD) is generally not observed for antiferromagnetic (AFM) states because XMCD signals from the antiparallelly coupled spins cancel each other. In this letter, we theoretically show the presence of an XMCD signal from compensated two-dimensional triangle AFM structures on a Kagome lattice. The calculation reveals the complete correspondence between the XMCD spectra and the sign of the spin chirality: the XMCD signal only appears when the spin chirality is negative. This XMCD signal originates from the different absorption coefficients of the three sublattices reflecting different charge density anisotropies and directions of spin and orbital magnetic moments.
\end{abstract}



\maketitle

X-ray polarization is used for the sensitive detection of the magnetic properties of materials. X-ray magnetic circular dichroism (XMCD) of x-ray absorption spectroscopy (XAS) is the most popular x-ray measurement technique used in studies of ferro- and ferri-magnetic states \cite{XMCD_Schutz, XMCD_Chen, XMCD_Thole, XMCD_Carra, XMCD_Tanaka, ferri_1}. XMCD in paramagnetic and antiferromagnetic (AFM) materials can be observed when a finite magnetization is induced by external magnetic fields or exchange coupling within the materials \cite{Au_XMCD, bilayer_AFM, Interface_AFM_Ohldag, Exchange_bias1, Exchangebias_Tsunoda, Exchange_bias2, Cr_AFM_XMCD,interface1}. One important question is whether magnetization is completely necessary for observing an XMCD signal. \

This appears to be true for most cases, but there are a few exceptions. Firstly, it is obvious in principle that a ferrimagnetic material at the compensation temperature, or with a composition with no magnetization, shows an XMCD signal when the ferrimagnetic coupling pair is composed of more than two different elements. Secondly, Sm$_{0.974}$Gd$_{0.026}$Al$_2$ \cite{SmAl_comp,SmAl_MCD} shows XMCD at the compensation temperature around 65 K due to cancellation between the spin and orbital magnetic moments at the Sm$^{3+}$ site. This study shows that the XMCD effect has the advantage of separating the spin and orbital components from the zero-magnetic moment. Although these zero-magnetic states are very special cases that can be experimentally realized by tuning the alloy composition and/or temperature very precisely, they demonstrate that an XMCD signal can in fact be observed even without a net magnetization. \

One of the most popular zero-magnetization states is an AFM ordered state with one magnetic element. It is expected that these AFM compounds do not exhibit XMCD effects because the spectral components of opposite spin states are considered to cancel out. However, Yamasaki $et \ al.$ recently propose an interesting conjecture that XMCD signals do exist for the AFM ordered state \cite{XMCD_Yamasaki}. In their study, an electron state, with $3z^2-r^2$ orbital in the triangular structure of the Kagome geometry, is employed to show the relationship between the spin chirality and the spin-dipole term, $T_{\rm z}$, which is included in the XMCD spin sum rule \cite{XMCD_Tanaka,XMCD_Oguchi}. They conclude that an XMCD signal is allowed when the total $T_{\rm z}$ composed in the triangular structure is not zero. However, since their study lacks a spectral calculation of XMCD, it may not fully present the correct interpretation of the XMCD effects for Kagome compounds. When the XMCD signal of the AFM structure is compared with the observed spectra, which usually includes artifacts or surface signals, the observations of the XMCD structure for the AFM ordered state can provide advantageous information for future experiments of such compounds. Moreover, the $T_{\rm z}$ dependence of the XMCD signal must be clarified for various types of electron configurations. For instance, because high-spin state $d^5$ systems have $|\vec{T}| = 0$, due to $|\vec{L}| = 0$ \cite{degroot}, the XMCD signal is expected to be absent, based on the arguments in ref. 17. To confirm this expectation, we performed the XMCD calculation for two divalent high-spin systems, Fe$^{2+}$ ($d^6$) and Mn$^{2+}$ ($d^5$), in triangular structure, and clarify the XMCD effect for the AFM ordered state, by considering the total spin and orbital moment for one of the magnetic elements. \

In this study, we consider a two-dimensional Kagome lattice, including a triangular structure and a crystal electric field (CEF), with a $D_{2h}$ point group, as shown in Fig. 1 (a). We employ a two-dimensional triangular structure and an orthorhombic-type CEF. In order to study various types of the AFM ordered states \cite{mtsuzuki}, we consider the molecular mean field (MF) of each site. Thus, the Hamiltonian is given by
\begin{equation}
H_{i}=H_{\rm atom}+H_{\rm CEF}+H_{\rm MF} ,
\end{equation}
Where index $i$ denotes a site of the triangular structure, shown by the dashed line in Fig. 1 (b). The first term $H_{\rm atom}$ is expressed as 
\begin{equation}
\begin{split}
H_{\rm atom}  \ = & \,\,\,\,\,\,\  \epsilon_{d}\sum_{\gamma}d^{\dagger}_{\gamma}d_{\gamma}+\zeta_{3d} \sum_{\gamma_{1},\gamma_{2}} (\bm{l\cdot s})_{\gamma_{1},\gamma_{2}} d^{\dagger}_{\gamma_{1}}d_{\gamma_{2}}    \\  
& +\epsilon_{p}\sum_{\gamma}p^{\dagger}_{\gamma}p_{\gamma} 
+\zeta_{2p} \sum_{\gamma_{1},\gamma_{2}} (\bm{l\cdot s})_{\gamma_{1},\gamma_{2}} p^{\dagger}_{\gamma_{1}}p_{\gamma_{2}}  \\           
& +\frac{1}{2} \sum_{\gamma_{1},\gamma_{2},\gamma_{3},\gamma_{4}} g_{dd} (\gamma_{1},\gamma_{2},\gamma_{3},\gamma_{4})\
d^{\dagger}_{\gamma_{1}}d_{\gamma_{2}}d^{\dagger}_{\gamma_{3}}d_{\gamma_{4}}   \\
&+\sum_{\gamma_{1},\gamma_{2},\gamma_{3},\gamma_{4}} g_{dp} (\gamma_{1},\gamma_{2},\gamma_{3},\gamma_{4}) \ d^{\dagger}_{\gamma_{1}}d_{\gamma_{2}}p^{\dagger}_{\gamma_{3}}p_{\gamma_{4}} , 
\end{split}
\end{equation}
where $d_{\gamma}^{\, \dagger}$ represents the creation operator for a 3$d$ electron state, including a combined index $\gamma$ with orbital and spin, and 
$p_{\gamma}^{\, \dagger}$ represents the creation operator for a 2$p$ core state. $H_{\rm atom}$ includes a 3$d$ level ($\epsilon_d$), spin-orbit coupling constant for the 3$d$ orbital ($\zeta_{3d}$), 2$p$ level ($\epsilon_p$), spin-orbit coupling constant for the 2$p$ orbital ($\zeta_{2p}$), coulomb interaction between the 3$d$ states ($g_{dd}$), and the Coulomb interaction between the 3$d$ and 2$p$ states ($g_{dp}$). These spin-orbit coupling constants and the Slater integral for $g_{dd}$ and $g_{dp}$ are estimated from the ionic calculation within the Hartree-Fock-Slater (HFS) method \cite{Cowan}. For the Slater integrals, 80$\%$ of the HFS values are used, as in previous studies \cite{Okada,Tanaka,Taguchi,Matsubara}. The second term $H_{\rm CEF}$ is determined by considering the one-electron potential of $D_{2h}$ symmetry, expressed as
\begin{equation}
\begin{split}
V_{\rm crys} \ = \,\,\,\,\, & B_{0}^{2}C_{0}^{(2)}+B_{2}^{2}(C_{2}^{(2)}+C_{-2}^{(2)})  \\ 
  +  &  B_{0}^{4}C_{0}^{(4)} +B_{2}^{4}(C_{2}^{(4)}+C_{-2}^{(4)}) +B_{4}^{4}(C_{4}^{(4)}+C_{-4}^{(4)}),
\end{split}    
\end{equation}
where
\begin{equation}
C_q^{(k)} = \sqrt{\frac{4\pi}{2k+1}}Y_{kq},
\end{equation}
and $Y_{kq}$ represents a spherical harmonic. 
Considering the previous study in Ref. 17, the CEF parameters (in eV) are $B^{2}_{0} = -0.7, B^{2}_{2} = 0.2, B^{4}_{0} = -2.7, B^{4}_{2} = 0.26$, and $B^{4}_{4} = 1.36$, and the $3z^2-r^2$ type
orbital is located in the lowest energy level; orbitals can be set to the lowest energy level by changing the CEF parameters \cite{CEF}.
The third term $H_{\rm MF}$ is expressed as 
\begin{equation}      
H_{\rm MF} = \sum_{\gamma_{1},\gamma_{2}} (\bm{h_{\rm MF}^{(i)}} \cdot \bm{s})_{\gamma_{1},\gamma_{2}} \ d^{\dagger}_{\gamma_{1}}d_{\gamma_{2}},
\end{equation}
where $h_{\rm MF}^{(i)}$ denotes the molecular field for the spin part of the $i$ site of the triangle geometry, surrounded by the dashed line in Fig. 1 (b). Thus, we employ the three-part Hamiltonian in eq. (1) to the red, blue, and orange targets, denoting three different spin states, as shown in Fig.1 (b).  \

\begin{figure}[!t]
\begin{center}
\includegraphics[width=90mm,clip]{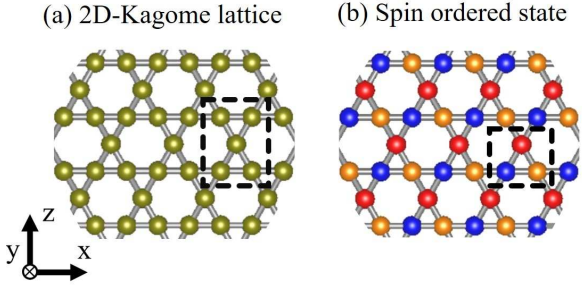}
\end{center}
\vspace{-2cm}
\caption{ (Color online) Crystal structure of a two-dimensional Kagome lattice. (a) The structure with a $D_{2h}$ point group is surrounded by a dashed line. (b) Diagram of the AFM state for a two-dimensional Kagome lattice. The red, blue, and orange targets have different spin directions, and the triangular lattice surrounded by the broken line is the unit cell of the magnetic structure.
}
\end{figure}

Figs. 2 and 3 show Fe 2$p$XAS and XMCD calculations for the AFM state of the triangular structure, considering Fe$^{2+}$; the upper and lower spectral structures show the XAS calculation ($\mu^+ + \mu^-$) and XMCD calculation ($\mu^+ - \mu^-$), respectively; $\mu^+$ and $\mu^-$ denote the absorption spectra recorded for x-ray photons with plus and minus helicity, respectively. We consider three types of AFM ordered state, with Figs. 2 (a), (b) and (c) showing the spin structure for Type-I, II, and III, respectively, and similarly for Fig. 3 (a-c). The difference between Figs. 2 and 3 is the direction of the incident x-rays, perpendicular to y-axis; in the former the direction is parallel to z-axis, and in the latter the direction is parallel to x-axis. In Figs. 2 and 3, the XAS structure of the red, blue, and orange targets is shown by the corresponding solid, colored lines, and the upper black line represents the summed XAS structure. Only the black line is observable in an actual experiment. Alternatively, the XMCD structure of the red, blue, and orange targets is given by the corresponding colored, dashed lines, with the lower black line representing the summed XMCD structure of the colored lines, again with only the black line being observable in an actual experiment.
\

\begin{figure}[!htb]
\begin{center}
\includegraphics[width=90mm,clip]{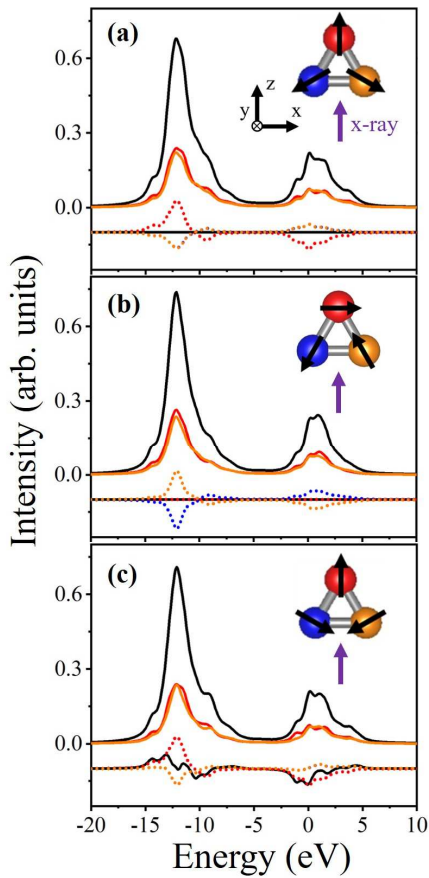}
\end{center}

\vspace{-1cm}
\caption{ (Color online) Fe 2$p$XAS and XMCD calculation for the AFM state of the triangular structure considering Fe$^{2+}$. The direction of incident x-ray is parallel to the z axis. The triangular structure is located in the xz plane. (a) The spin direction of the red target is parallel to the z axis, and the blue and orange targets are rotated 120- and 240-degree  clockwise from the red, respectively. (b) The spin direction of the red target is parallel to the x axis, and the blue and orange targets are rotated 240- and 120-degree clockwise from the red, respectively. (c) The spin direction of the red target is parallel to the z axis, and the rotation of the blue and orange targets has the same conditions as in Fig. 2(b).
}
\end{figure}

\begin{figure}[!htb]
\begin{center}
\includegraphics[width=90mm,clip]{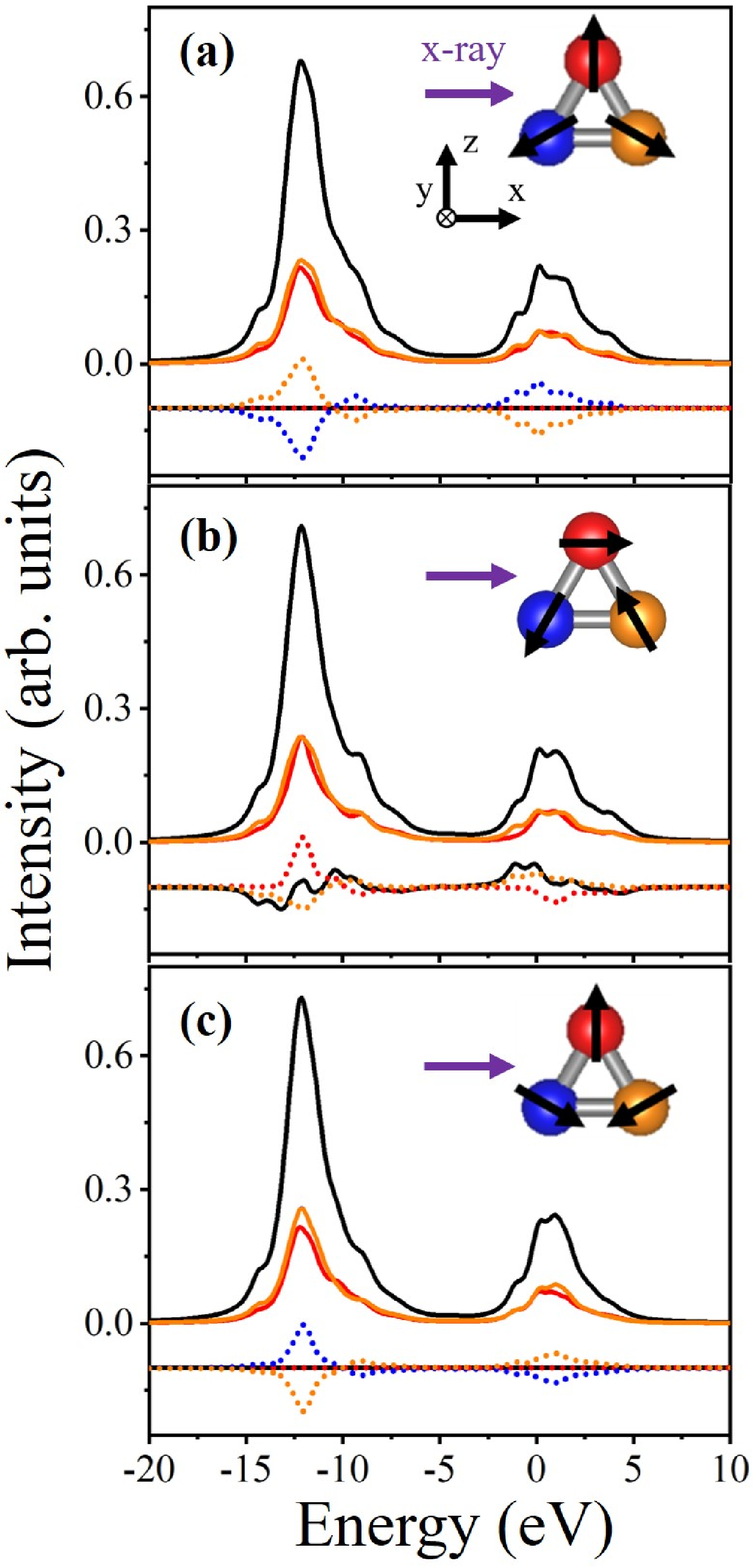}
\abovecaptionskip=-15pt
\end{center}

\vspace{-1cm}
\caption{ (Color online)  Fe 2$p$XAS and XMCD calculation for the AFM state of the triangular structure considering Fe$^{2+}$. The direction of incident x-ray is parallel to the x axis. The triangular structure is located in the xz plane. (a) The spin direction of the red target is parallel to the z axis, and the blue and orange targets are rotated 120- and 240- degree clockwise from the red, respectively. (b) The spin direction of the red target is parallel to the x axis, and the blue and orange targets are 240- and 120- degree rotated clockwise form the red, respectively. (c) The spin direction of the red target is parallel to the z axis, and the rotation of the blue and orange targets is the same conditions as in Fig. 3(b).
}
\end{figure}

We see that the XMCD signal of the Type-I AFM structure does not arise in the geometries shown in Figs. 2 (a) and 3 (a), whereas the Type-II and III AFM structures do show an XMCD signal in Figs.3 (b) and 2 (c), respectively. We find that the XMCD signal depends on the spin chirality in the triangular structure, because the Type-I AFM structure shows positive chirality and the Type-II and III AFM structures show negative chirality, where the rotational direction of the spin component for positive chirality is opposite to that of negative chirality. Moreover, the XMCD signal of the negative chirality AFM structure is dependent on the incident x-ray direction, which is consistent with the previous report \cite{XMCD_Yamasaki}. The XMCD signal for $d^6$ system is essential for the total $T_{\rm z}$ term because the composed total spin magnetic moment  $|\vec{S}^{(\rm{total})}| = |\vec{S}^{\rm{(red)}} + \vec{S}^{\rm{(blue)}} + \vec{S}^{\rm(orange)} | = 0$ and orbital moment $|\vec{L}^{(\rm{total})}| = |\vec{L}^{\rm{(red)}} + \vec{L}^{\rm{(blue)}} + \vec{L}^{\rm(orange)} | = 0$ of the triangular lattice.

\begin{figure}[!htb]
\begin{center}
\includegraphics[width=90mm,clip]{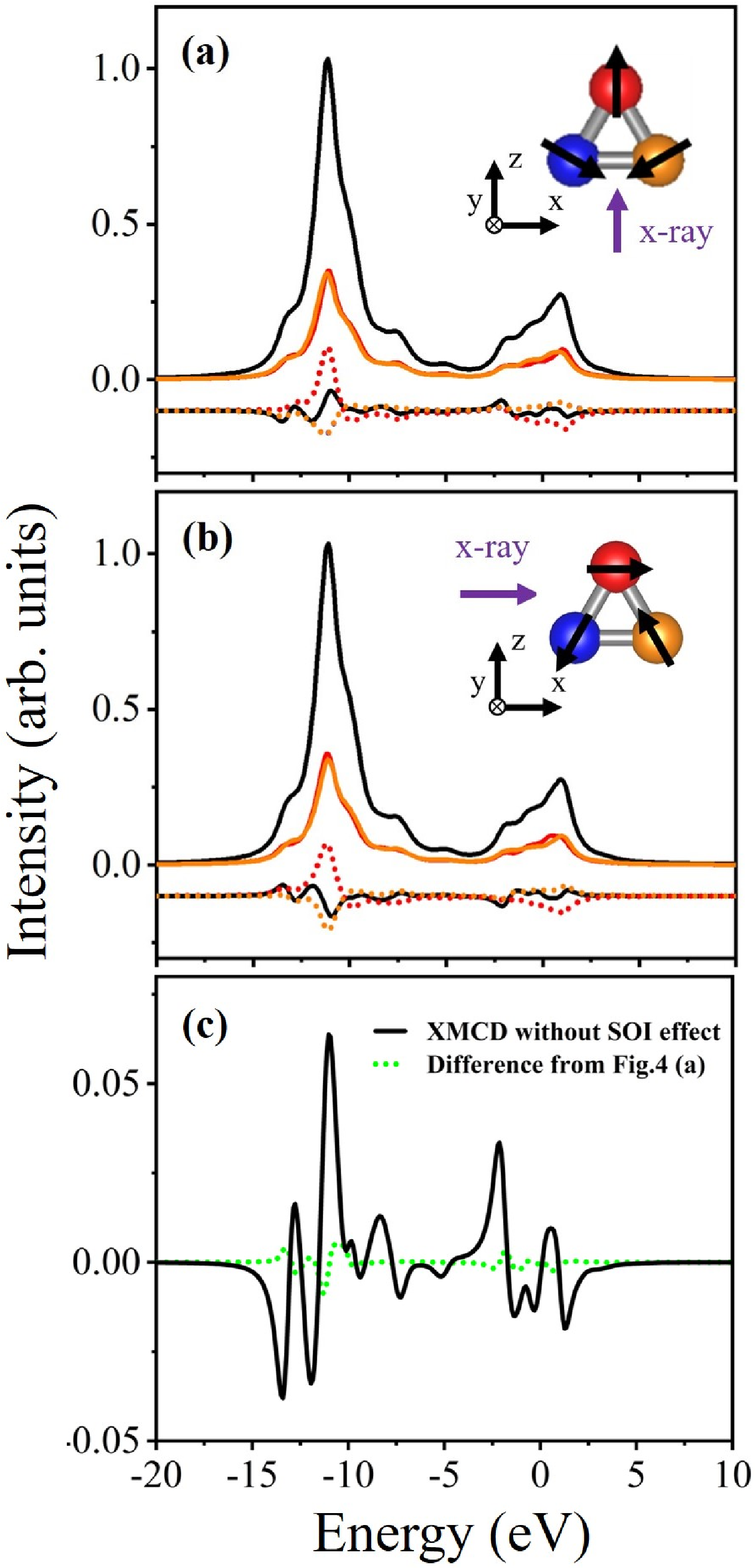}
\end{center}
\caption{ (Color online)  Mn 2$p$XAS and MCD calculation for the AFM state of the triangular structure considering Mn$^{2+}$. The triangular grating geometry is located in the xz plane. (a) The spin direction of the red target is parallel to the z axis, and the blue and orange targets are rotated 240- and 120-degree clockwise from the red, respectively. The direction of incident x-ray is parallel to the z axis. (b) The spin direction of the red target is parallel to the x axis, and the rotations of the blue and orange targets are same as in Fig. 4(a). The direction of incident x-ray is parallel to the x axis. 
(c) Spin-orbital interaction effect; the spin ordered state and incident photon settings are the same as in Fig.4 (a). the black solid line shows XMCD calculation considering $\zeta_{3d}$ = 0, and the green dotted line is the difference of XMCD signal (black line) in Fig.4 (a) and (c).
}
\end{figure}  
 
 Next, to study the behavior of the total orbital angular momentum in the XMCD for the AFM Kagome structure, we examine the Mn 2$p$XAS and XMCD calculations of the negative spin chirality structure for Mn$^{2+}$. In Figs. 4 (a) and (b), the upper spectral structure shows the XAS calculation, with Fig. 4 (a) having the same conditions for the spin structure and incident x-ray direction as in Fig. 2 (c). Similarly, the lower spectral structure in Figs. 4 (a) and (b) shows the calculation for XMCD, where Fig. 4 (b) has the same condition for spin structure and incident x-ray direction as in Fig. 3 (b). Subsequently, we find that the negative spin chirality AFM structure has an XMCD signal. Notably, the Type-I AFM structure of Mn$^{2+}$ does not have an XMCD signal, and the negative spin chirality structure of Mn$^{2+}$ depends on the incident x-ray direction for the appearance of an XMCD signal. 
Thus, the calculated results for the XMCD spectra are qualitatively very similar between those for Mn$^{2+}$ and Fe$^{2+}$. However, there are major differences in their orbital angular momentum, on which $T_{\rm z}$ depends. The expectation values of the total orbital moment, $|\vec{L}|$, for Mn$^{2+}$ and Fe$^{2+}$ were calculated to be 0.0028 and 2.0013, respectively, when their spin-orbit interactions were considered as were in Fig.4 (a) and (b). In order to investigate a more fundamental issue, we also calculate XMCD spectrum for Mn$^{2+}$ without the spin-orbit interaction and show the result in Fig.4(c). In Fig.4(c), the obtained XMCD effect is significant in spite of $|\vec{L}|$=0 in Mn$^{2+}$ and is almost identical to the case with the spin-orbit interaction. This result clearly indicates that $T_{\rm z}$ is not essential for the XMCD effect originating from the AFM Kagome structure.
\ 

Taking the Mn$^{2+}$ case into account, the XMCD signal for the AFM Kagome structure must be explained without using the $T_{\rm z}$ term. 
To explain the origin of this XMCD signal, we consider one electron state at three targets in the triangular structure with positive and negative spin chiralities, as shown in Fig. 5 (a). Fig. 5 (b) shows the geometric layout between the triangle structure and the incident x-ray directions, with the solid arrow corresponding to Figs. 2 (a-c) and 4 (a) and the dashed arrow corresponding to Figs. 3 (a-c) and 4 (b). In Fig. 5 (a), the direction of spin at each target is expressed by an arrow on a d orbital. Here, we illustrate the $3z^2-r^2$ type orbital as a realistic example. The long direction of the $3z^2-r^2$ orbital changes by 120 degrees when the position of the triangle is transferred clockwise; this restriction is determined by the symmetry of the crystal structure. Conversely, depending on the sign of the spin chirality, the direction of spin is changed by 120 or -120 degrees for positive or negative spin chirality, respectively. As shown at the bottom of Fig. 5 (a), considering the common local quantization axis, the relative configuration between the spin and the anisotropic charge distribution is equivalent for positive spin chirality. On the other hand, when the spin chirality is negative, the three electronic states in each sublattice have different characteristics. Arranging the x-ray direction under the same local quantization axis as shown on the left side of Fig. 5 (b), the understanding of the unconventional XMCD effect is improved. For positive spin chirality, the XMCD contributions from each sublattice cancel each other owing to the equivalency of the electronic states under the restricted geometric layout in Fig. 5 (b).  Thus, the complete XMCD signal does not arise as in Figs. 2 (a) and 3 (a). Then, for the negative spin chirality, the three different electronic states have a different absorption coefficient of the XAS process, and thus the XMCD signal arises as seen in Figs. 2 (c) and 3 (b). Moreover, we focus on the spectral shapes of the XMCD signal. As regards the spectral shapes of the conventional XMCD signal, the strongest XMCD signal often arises in the main peak of the XAS, as shown by the colored lines in the figures. On the other hand, for the negative chirality case, we see obvious peaks, not in the main peak, but in the XAS multiplet structure, as shown by the black lines. Therefore, we consider that the XMCD signal of negative spin chirality has a characteristic structure caused by the different absorption coefficient of the three electronic states.

\begin{figure}[!htb]
\begin{center}
\includegraphics[width=90mm,clip]{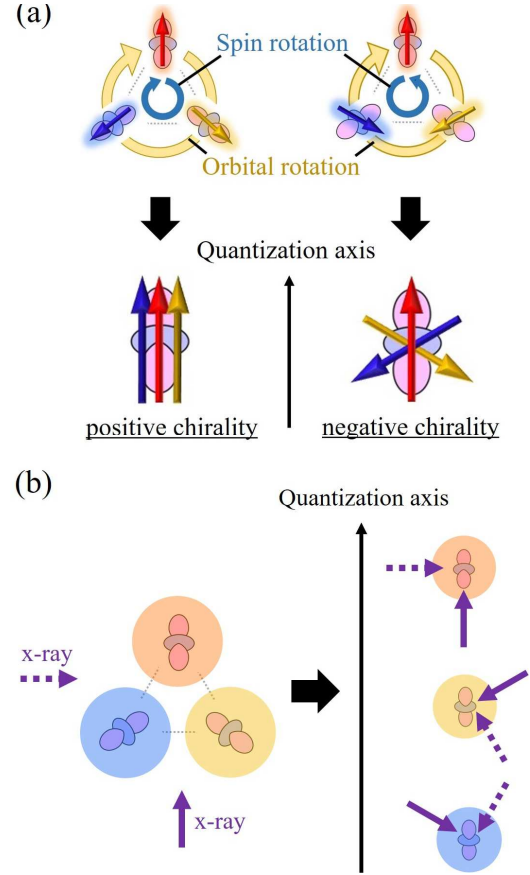}
\end{center}
\vspace{-1mm}
\caption{ (Color online) (a) A state of positive chirality (left) and negative chirality (right) of the triangular geometry in the upper half. When one target is transfered clockwise to the neighbor position, the spin and charge distribution changes by 120 degree following the spin and orbital rotation directions, respectively. In the bottom half, the spin and charge distributions are arranged by considering one quantization axis. (b) Geometry between incident x-ray direction and triangular structure is described in the left side; the solid arrow corresponds to Figs. 2 (a-c) and Fig. 4 (a) and the dotted arrow corresponds to Figs. 3 (a-c) and Fig. 4 (b). By arranging the incident x-ray direction using the above quantization axis, the incident x-ray direction of each target is shown in the right side.
}
\end{figure}

Finally, we compare our study with the previous study in Ref. 17. The previous study shows the relationship between the XMCD signal and the total spin dipole operator, and the origin of the XMCD signal is the uncompensated $T_{\rm z}$ term in the spin structure with negative spin chirality. In contrast, we have demonstrated here that the XMCD signal is clearly observed even for $T_{\rm z}$ = 0. From our simple discussion above, we attributed the origin of this XMCD signal to non-equivalent electronic states at each sublattice. Namely, we provide a general view of the XMCD effect for kagome AFM compounds, and the results of our study also mean that sublattice-dependent non-equivalent electronic states in cluster multipole states other than Kagome AFM could also be observable by XMCD. In future work, such a simple description (as in Fig. 5) will be helpful in determining the presence of XMCD in characteristic AFM materials before experimentation.
\ 

In summary, our calculation demonstrates the XMCD spectra for the AFM state of a two-dimensional triangular structure, including either Fe$^{2+}$ and Mn$^{2+}$ ions. Our numerical calculations show that the sign of spin chirality is determined using the XMCD along with the incident x-ray direction, which is consistent with the previous report \cite{XMCD_Yamasaki}. Our results reveal not only the details of the XMCD features for a Kagome AFM structure, by calculating the three XMCD signals on the triangular structure, but also the general response of XMCD on the AFM state constructed by one element, by showing the spectral features for the zero $T_{\rm z}$ case of a $d^5$ system \cite{d5}. This means that XMCD has an advantage in detecting different electronic states of the same element, and the XMCD signal under AFM condition appears in structures other than just the Kagome structure. We believe that XMCD is useful for investigating complicated and non-trivial spin structures, for example, quantum frustration coupled with time-reversal-symmetry-broken AFM structures \cite{mtsuzuki}. Our work contributes significantly to the development of devices including non-trivial spin structure, such as high density non-volatile memory with no restriction of the magnetic interference between storage cells \cite{nature,nature2}.\

\begin{acknowledgments}
The authors are grateful to Y. Yamasaki, H. Nakao, D. Billington, M. Mizumaki, A. Yasui, and T. Uozumi for helpful discussions.
\end{acknowledgments}

\bibliography{basename of .bib file}

\begin{thebibliography}{9}
\bibitem{CEF} It is noted that a specific target material was not employed in this study. When the target material is chosen, we need to estimate the value of the CEF splitting using first-principles calculations.
\bibitem{d5} This case of the $d^5$ system is ideal due to no advantage of energy stabilization where 
ferro-octopole ordering with a finite expectation value of the spin-dipole operator $T_{\rm z}$ arises. However, in order to clarify a universal principle, we employ the $d^5$ system.
\end{thebibliography}

\end{document}